\documentclass{emulateapj}
\slugcomment{DRAFT \today}
\shorttitle{Angular Momentum of Protostars}
\shortauthors{Covey, Greene, Doppmann \& Lada}

\begin{document}

\title{The Angular Momentum Content and Evolution of Class I and Flat-Spectrum Protostars}

\author{Kevin~R.~Covey\altaffilmark{2}, Thomas P. Greene\altaffilmark{3}, Greg W. Doppmann\altaffilmark{3,4}, Charles J. Lada\altaffilmark{5}}

\altaffiltext{2}{University of Washington, Department of Astronomy, Box 351580, Seattle, WA 98195; covey@astro.washington.edu}
\altaffiltext{3}{NASA Ames Research Center, Mail Stop 245-6, Moffett Field, CA 98035-1000; thomas.p.greene@nasa.gov}
\altaffiltext{4}{Present Address: Gemini Observatory, Southern Operations Center, Association of Universities for Research in Astronomy, Inc., Casilla 603, La Serena, Chile; doppmann@gemini.edu}
\altaffiltext{5}{Harvard-Smithsonian Center for Astrophysics, 60 Garden Street, Cambridge, MA 02138; clada@cfa.harvard.edu}

%email address: 
%covey@astro.washington.edu

\begin{abstract}
We report on the angular momentum content of heavily embedded protostars based on our analysis of the projected rotation velocities ($v$ sin $i$s) of 38 Class I/flat spectrum young stellar objects recently presented by \citet{Doppmann2005}\footnote{The data presented herein were obtained at the W.M. Keck Observatory, which is operated as a scientific partnership among the California Institute of Technology, the University of California and the National Aeronautics and Space Administration. The Observatory was made possible by the generous financial support of the W.M. Keck Foundation.}.  After correcting for projection effects, we find that infrared-selected Class I/flat spectrum objects rotate significantly more quickly (median equatorial rotation velocity $\sim$ 38 km/sec) than Classical T Tauri stars (CTTSs; median equatorial rotation velocity $\sim$ 18 km/sec) in the $\rho$ Ophiuchi and Taurus-Aurigae regions.  Projected rotation velocity ($v$ sin $i$) is weakly correlated with T$_{eff}$ in our sample, but does not seem to correlate with Br $\gamma$ emission (a common accretion tracer), the amount of excess continuum veiling (r$_k$), or the slope of the SED between the near and mid IR ($\alpha$).  The detected difference in rotation speeds between Class I/flat spectrum sources and CTTSs proves difficult to explain without some mechanism which transfers angular momentum out of the protostar between the two phases.  Assuming Class I/flat spectrum sources possess physical characteristics (M$_*$,R$_*$,B$_*$) typical of pre-main sequence stars, fully disk locked Class I objects should have co-rotation radii within their protostellar disks that match well (within 30\%) with the predicted magnetic coupling radii of \citet{Shu1994}.  The factor of two difference in rotation rates between Class I/flat spectrum and CTTS sources, when interpreted in the context of disk locking models, also imply a factor of 5 or greater difference in mass accretion rates between the two phases.  A lower limit of \.M $\sim 10^{-8}$ M$_{\odot}$/year for objects transitioning from the Class I/flat spectrum stage to CTTSs is required to account for the difference in rotation rates of the two classes by angular momentum extraction through a viscous disk via magnetic coupling. 
\end{abstract}


\section{Introduction}

Over the past two decades evidence has accumulated that the angular momentum content of low mass stars undergo significant modification during their first 100 million years.  This evolution has been best documented for cluster stars which have evolved to the T Tauri phase and beyond \citep[for informative reviews, see:][]{Bodenheimer1995,Mathieu2003}.  A key discovery is that the majority of low mass stars in the classical T Tauri phase are relatively slow rotators ($v$ sin $i <$ 20 km s$^{-1}$; Bouvier et al. 1986, Hartmann et al. 1986, Hartmann \& Stauffer 1989, Bouvier 1990) contradicting expectations that accretion of high specific angular momentum material from a circumstellar disk will drive stars to rotate near breakup velocity \citep{Hartmann1989}.  Additionally, while the peak rotation velocity for a given population of young low mass stars consistently occurs at low velocities, a high velocity tail of rapid rotators ($v$ sin $i >$ 50 km s$^{-1}$) grows as a fraction of the total cluster population until an age of $\sim$ 100 Myrs, based on observations of young clusters such as IC 2602 \& 2391 \citep{Stauffer1997,Barnes1999}, $\alpha$ Per \citep{Stauffer1989,Prosser1992,Prosser1994} and the Pleiades \citep{Stauffer1984,Stauffer1987,Soderblom1993,Jones1996,Queloz1998}. 

The observational record is harder to interpret at the youngest ages.  Investigations of Pre-Main Sequence (PMS) stars in Orion with photometric rotation periods have revealed a bimodal period distribution for higher mass (M $>$ 0.25 M$_{\odot}$ according to the models of D'Antona \& Mazzitelli 1994, M $\gtrsim$ 0.5 M$_{\odot}$ according to the models of Baraffe 1998) stars, but only a single peak for lower mass stars \citep{Herbst2000,Herbst2001}.  However, a similar study of Orion PMS stars \citep{Rebull2001} was unable to detect this characteristic shift in the relation between rotation and mass.  Likewise, early investigators found a link between rotation rates of T Tauri stars in Tau-Aur and Orion and infrared excesses (indicative of the presence of a circumstellar disk) such that slow rotators possesed large excesses, which rapid rotators lacked \citep{Edwards1993}.  These findings have been interpreted in support of a paradigm in which T Tauri stars with large circumstellar disks (Classical T Tauri Stars [CTTSs], or Class II objects in the classification system of Lada 1987) are forced to rotate slowly via star-disk interactions, while T Tauri stars whose disks have dissipated (Weak-lined T Tauri Stars [WTTSs], or Class III objects) spin up to become rapid rotators as they contract onto the main sequence.  More recent investigations in Orion, though, have found no such strong correlations between rotation rates and photometric infrared excesses of T Tauri stars \citep{Stassun1999,Rebull2001,Rhode2001}.  Indeed, it has even been suggested that the form of the correlation between rotation rates and infrared excesses that were reported in the past are contrary to what should be expected from theoretical models of angular momentum transfer \citep{Stassun2001}.  Part of this confusion may stem from difficulties in using photometry alone to assess the presence or absence of disk material around T Tauri Stars; new work by \citet{Sicilia-Aguilar2005} using high resolution (R $\sim$ 35,000) spectroscopy to classify objects as either CTTSs or WTTSs on the basis of emission line widths suggests that photometric criteria for infrared excesses do not perfectly segregrate objects on the basis of their accretion strengths.  \citet{Sicilia-Aguilar2005} find that the use of spectroscopic criteria to separate CTTSs from WTTSs results in a clear difference between the rotation properties of the two types of stars, where WTTSs appear to rotate significantly more rapidly than CTTSs. 

Even less is known about stars in earlier evolutionary stages than T Tauri stars.  To date, only a handful of embedded Class I and flat spectrum protostars have rotation rates measured from analysis of infrared photospheric spectral features \citep{Greene1997,Greene2000,Greene2002}.  These objects appear to have faster rotation velocities ($v$ sin $i \sim$ 40 km s$^{-1}$) than their more visible T Tauri successors.  In the optical, \citet{White2004} (hereafter WH04) recently analyzed photospheric spectral features for a sample of 11 less heavily embedded (and thus optically visible) Class I sources in Taurus and found slower rotation velocities ($v$ sin $i \sim$ 20 km s$^{-1}$) that were comparable to those detected in a companion sample of Class II Herbig-Haro drivers.  It thus appears that Class I protostars can exhibit a considerable range of rotation velocities (10-50 km s$^{-1}$) whose magnitude may be related to how deeply embedded the objects are.  The lack of a single large, homogenous sample, however, has as yet prevented strong statistical constraints on the angular momentum distribution of Class I protostars.

The observational difficulty (due to the high levels of extinction and significant veiling emission from circumstellar material) of measuring rotation rates for Class I protostars is particularly unfortunate, as it appears a great deal of angular momentum evolution may occur during the protostellar accretion phase.  The precursor angular momentum content of low mass stars can be assessed by observations of molecular cores within star formation regions, believed to be evolutionary predecessors to more evolved, collapsing protostars.  Observations of these cores in molecular transitions which trace only the densest gas (e.g. NH$_3$; Goodman et al. 1993, Barranco \& Goodman 1998) reveal linear velocity gradients expected for objects in solid body rotation, with a resultant estimated specific angular momentum of a typical dense core of 10$^{21}$ cm$^2$ s$^{-1}$.   A comparison to the estimated specific angular momentum for isolated T Tauri stars of 10$^{17}$ cm$^2$ s$^{-1}$ \citep{Hartmann1986} reveals that protostars must lose 99.99$\%$ or more of their angular momentum prior to entering the T Tauri stage.  

A number of theoretical mechanisms for transporting angular momentum from an accreting protostar have been suggested.  Disk locking models \citep{Konigl1991,Shu1994} envision a stellar magnetic field frozen into the midplane of a circumstellar disk at a point where the angular (Keplerian) rotation velocity is slightly lower than the stellar rotation velocity.  This causes the star to be spun down by magnetic drag while the specific angular momentum of the disk material increases, later being dissipated by viscous transport or disk winds.  Wind models which do not invoke disk locking \citep{Tout1992} involve a transfer of angular momentum from the star to circumstellar material through ionized stellar winds which are constrained to flow along rotating magnetic field lines.  Other mechanisms, such as dynamical interactions with gravitational instabilities in protostellar disks \citep{Durisen2003} or fragmentation of the core and disk into a multiple star system, also allow transfer of angular momentum away from the central star.  

Direct observational tests to distinguish between these angular momentum transport models have been difficult, particularly insofar as much of the evolution has already taken place by the time a protostar emerges from its accretion envelope as a classical T Tauri star (CTTS).  In an attempt to address this lack of observational data concerning the rotational properties of younger, more heavily accreting protostars, we present here an analysis of rotational velocities derived from a spectroscopic survey of Class I and flat spectrum protostars in the Tau-Aur, $\rho$ Oph, Ser and other star forming regions presented by \citet{Doppmann2005}.  \S 2 describes briefly the observations and projected rotation velocity ($v$ sin $i$) determinations of the Class I and flat spectrum sample, as well as the creation of a comparison sample of T Tauri stars in Tau-Aur and $\rho$ Oph with measured $v$ sin $i$ values.  \S 3 explains the implementation of a method for numerically inverting the observed $v$ sin $i$ distribution to reveal the most likely equatorial rotation velocity distribution.  In \S 4 we compare the rotation of the Class I and flat spectrum sample to that of T Tauri stars, finding that the two samples are different to a high degree of statistical confidence.  We analyze the implications of those differences in the context of various models of angular momentum transport for evolving protostars in \S 5, with \S 6 summarizing our main conclusions.

\section{Observations \& Projected Rotation Velocity Determinations}

\subsection{Class I and flat spectrum sources}

The observed protostellar rotation velocities used in this study were derived from our larger Class I and flat spectrum survey described in full by \citet{Doppmann2005} (hereafter D05).  We summarize here those details of the observations and data analysis of the D05 survey most relevant for this work.  Spectra of Class I and flat spectrum (hereafter CI/FS) sources were observed over the course of 13 nights between May 2000 and June 2003 using the near infrared echelle spectrograph NIRSPEC \citep{mclean1998} on the 10 m Keck 2 telescope.  Spectra were acquired with a 0.58'' (4 pixel) wide slit (allowing R $\sim$ 18,000) utilizing a 1024x1024 pixel InSb detector array with the NIRSPEC-7 blocking filter under median $\sim$ 0.6'' seeing.  The  NIRSPEC gratings were configured such that 4 prominent sets of spectral lines could be simultaneously observed in non-continuous orders: 1) the 2.1066 $\mu$m Mg and 2.1099 $\mu$m Al lines, 2) the 2.1661 $\mu$m Br $\gamma$ line, 3) the 2.2062,2.2090 $\mu$m Na doublet and the 2.2233 $\mu$m H$_{2}$ line, and 4) the 2.2935 $\mu$m CO bandhead.  

 All spectra were flat fielded, cosmic ray cleaned, sky subtracted, extracted, wavelength calibrated and corrected for telluric absorption using standard IRAF packages and procedures \citep{Massey1992,Massey1997}.  The observed NIRSPEC spectra of each protostar was then matched to a grid of synthetic spectra, constructed by MOOG \citep{Sneden1973} from the NEXTGEN model atmospheres \citep{Hauschildt1999} and modified to represent various veilings, rotation velocities and radial velocities.  Artificial rotation broadening of synthetic spectra was accomplished with the use of a rotation broadening profile with an assumed limb-darkening coefficient of 0.6 \citep{Gray1992}.  Least squares minimization of the residuals between the observed spectra and synthetic spectral grid then produced estimates of the effective temperature (T$_{eff}$), surface gravity (log \textit{g}), radial velocity (v$_{r}$), projected rotation velocity ($v$ sin $i$) and veiling (r$_{k}$) for each protostar.  The spectral resolution of the observations translates to a velocity resolution of $\sim$ 14 km s$^{-1}$; detected rotational velocities were calculated with this slit broadening removed in quadrature.  Error estimates for the derived rotational velocities were calculated from the shape of the local minimum in the fitting residual space sampled across the rotation velocity dimension and were typically 3-4 km s$^{-1}$.  See D05 for individual $v$ sin $i$ velocities and uncertainties. 

To confirm the accuracy of our derived $v$ sin $i$ estimates, we have compared measurements of the rotation velocities of a number of our targets with values previously reported in the literature for those objects.  The derived $v$ sin $i$s for two spectral type standards in our sample (Gliese 791.2 and 1245a) were compared with the projected rotation velocities derived for the same objects by \citet{Mohanty2003}.  We find rotation velocities for these stars of 34.3$\pm$1.3 and 22.5$\pm$2.2 km s$^{-1}$ respectively, consistent within the errors with the values found by \citet{Mohanty2003} of 32$\pm$2 and 22.5$\pm$2 km s$^{-1}$.  WH04 recently used optical spectra to produce $v$ sin $i$ estimates for 5 of the targets involved in this study: of those 5 shared targets, 4 (IRAS 04158+2805, DG Tau, GV Tau S and IRAS 04489+3042) produced derived velocities which matched within the errors on the measurements.  WH04 also record an upper limit of 15 km s$^{-1}$ for IRAS 04016+2610, while the measured $v$ sin $i$ value we find for IRAS 04016+2610 is significantly larger: $v$ sin $i =$ 46$\pm$2.7 km s$^{-1}$.  WH04, however, flag their derived $v$ sin $i$ value for IRAS 04016+2610 as highly uncertain due to the extremely low signal to noise of their sprectra; as our spectrum is extremely well measured (S/N = 240) we assume the value reported by WH04 for this particular object is in error due to the faintness of its optical detection.  Our measured $v$ sin $i$ value of 24$\pm$2.0 km s$^{-1}$ for DG Tau (an optically visible T Tauri star with a flat SED) also matches well with the previously determined value of 21.7$\pm$6.3 km s$^{-1}$ found by \citet{Hartmann1989}.  

Additionally, most field stars later than spectral type G have rotation velocities below the detection limit of this survey.  As we would therefore expect, the 26 slowly rotating spectral type standards observed by D05 (all later than G7) provided only one with a significant ($v >$ 14 km s$^{-1}$) detection of rotation velocity: Gliese 569B.  With a measured $v$ sin $i$ of 20 km s$^{-1}$, Gliese 569B has also been detected through adaptive optics imaging to be a binary system \citep{Martin2000} -- it is possible the broader linewidth of the composite spectrum is due more to combining spectra of individual components at different orbital velocities than any intrinsic stellar rotation.  The high level of agreement between the values reported by \citet{White2004}, \citet{Mohanty2003}, \citet{Hartmann1989} and our derived $v$ sin $i$ values for shared targets, as well as our upper limits for 25 of 26 late type field stars, support the accuracy of both our derived rotational velocities and our corresponding error estimates.

The full CI/FS sample of D05 consists of 41 objects with measurable absorption lines.  Some care, however, must be applied before blindly accepting all such objects as input data for a study of this nature.  Spectra of objects which are actually unresolved binaries could cause the characteristic rotational velocity derived here for protostars to be an overestimate, as the blending of spectral lines from two binary components at slightly different radial velocities could be misinterpreted as rotational broadening of a single line in one object.  To attempt to reduce this error, we have taken the following two steps: a) we have removed from our sample those objects known to be in binaries with separations on the order of or less than the slit width with which the observations were taken \citep[WL1 and Haro 6-28;][]{Haisch2004,Simon1995}, and b) we have cross correlated our target spectra with similar spectral standards using the IRAF task FXCOR to see if any sources show structure in their cross correlation results across multiple orders which suggests an observed spectra may be the sum of two individual components.  This additional screening process identified only one object with signs of unresolved binarity; this source (IRS51) has been eliminated from our sample as well.  An additional sign of binarity would be sources whose radial velocities are discrepant from the mean cluster velocity due to orbital motions.  In our separate work investigating the radial velocity distribution of these sources \citep{Covey2005b} we have identified four objects within the CI/FS sample with radial velocities that are 3$\sigma$ outliers with respect to the velocity of the local CO gas.  Of these four sources, two are relatively slow rotators ($v$ sin $i <$ 30 km s$^{-1}$) while two are more rapid rotators ($v$ sin $i >$ 40 km s$^{-1}$).  We discuss more fully their possible binary status in our work on radial velocities, but with no concrete proof of their binarity, have left them in the rotation sample.  Therefore, to the fullest extent possible we have tested for and eliminated unresolved binaries from our sample, but we welcome future adaptive optics observations of these systems to place tighter limits on the possible binary companions. 

Having been culled of possible binaries, the remaining 38 sources from D05 come from several nearby star formation regions, most notably Taurus Aurigae (10 targets), $\rho$ Ophiuchi (14 targets) and Serpens (9 targets).  5 additional targets were selected from R CrA and Perseus.  An additional 11 targets were observed in the course of the D05 survey but proved to have only smooth continuum emission with no photospheric absorption lines in the spectral regions studied.  All targets were chosen from surveys of local star formation regions utilizing IRAS, ISO, or ground based observations to assess the slope of member SEDs from the near to mid IR \citep{Wilking1989,Kenyon1990,Greene1994,Kenyon1995,Bontemps2001,Kaas2004}.  The sample was limited to objects with K $<$ 11 in general, with some additional preference for objects with low bolometric luminosities to aggressively sample the low end of the mass spectrum.  The top panel of Figure 1 displays a histogram of the raw $v$ sin $i$ velocities observed for the 38 protostars in the sample. 

\begin{figure}
\plotone{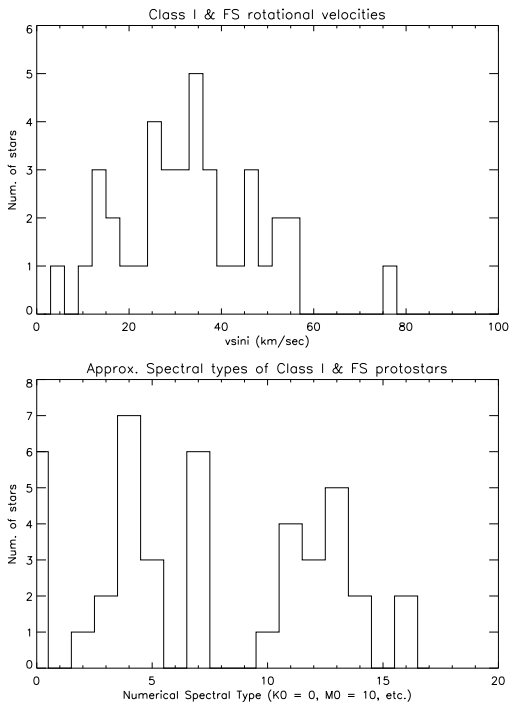} 
\caption{\scriptsize{Upper panel: A histogram, with 3 km/sec bins, of the observed $v$ sin $i$ values for a sample of 38 class I and flat spectrum protostars observed by D05.  Lower panel: The distribution of spectral types for the protostars in the Class I and flat spectrum sample, based on the derived T$_{eff}$ of the object and the dwarf T$_{eff}$-spectral type conversion presented by \citet{deJager1987}.}}
\label{fig1}
\end{figure}

To ensure the validity of conclusions concerning protostellar evolution derived from comparisons between the protostellar sample considered here and a sample of T Tauri stars (described in \S 2.2 below), it is important to ensure to the maximum extent possible that the two samples contain objects with similar distributions of final main sequence stellar masses.  Pre-main sequence stars largely contract along Hayashi tracks towards the low end of the main sequence, so the evolution of a low mass star from the CI/FS phase to the T Tauri phase should not introduce large changes in its effective temperature or spectral type \textit{in the absence of accretion.}  In order to aid the comparison of the rotation properties of our CI/FSs to those of a sample of T Tauri stars, we have calculated the approximate spectral type for each of the protostars in our sample based on the derived T$_{eff}$ of the object and the dwarf T$_{eff}$-spectral type conversion presented by \citet{deJager1987}. In the range of T$_{eff}$s relevant to this work, dwarf and PMS T$_{eff}$/spectral type relations differ only by a few hundred degrees K.  Given the scale of other uncertainties in the process (see next paragraph) and that many T Tauri stars have been spectral typed using dwarf indices, we have chosen to perform our T$_{eff}$/spectral type conversion with respect to a dwarf scale.  The resultant approximate spectral type distribution of the Class I and flat spectrum sample is presented in the bottom panel of Figure 1.

CI/FS sources do, however, have derived mass accretion rates \.M $\sim 10^{-6}$ M$_{\odot}$ yr$^{-1}$ \citep{Wilking1989,Greene2002}; over an assumed lifetime of the CI/FS phase of 10$^5$ yrs, this would amount to an additional maximum of 0.1 M$_{\odot}$ being added to the source.  Variations of 0.1 M$_{\odot}$ in the mass of main sequence stars with spectral types typical of our CI/FS sample ($\sim$ K7) correspond to variations in spectral type of $\sim$ 3 spectral type subclasses.  Therefore, the unknown future accretion histories of CI/FS objects do introduce some uncertainty into our ability to use spectral type or T$_{eff}$ as a proxy for the final main sequence mass of objects in our CI/FS sample.  However, the scale of the resulting bias does not seem large enough to create a significant mismatch between the final main sequence masses of CI/FS and T Tauri samples with identicial current spectral type ranges.  We thus conclude that though more detailed knowledge of the future accretion history of our CI/FS sample would be desirable to properly construct an analogous sample of T Tauri stars, constructing samples with similar current spectral type distributions allow a reasonable investigation of evolutionary effects.

\subsection{T Tauri stars}

In order to facilitate the interpretation of the measured $v$ sin $i$ distribution of Class I and flat spectrum sources in the context of Pre-Main Sequence (PMS) evolution, we have assembled a sample of $\sim$ 150 pre-main sequence stars with spectroscopically measured $v$ sin $i$ values from the literature \citep{Rebull2004,Doppmann2003B,White2003,Jayawardhana2003a,Greene2000,Clarke2000,Wichmann2000,Luhman1999,Greene1997,Preibisch1997,Padgett1996,Kenyon1995,Neuhaeuser1995,Gameiro1993,Magazzu1992,Bouvier1990,Hartmann1989,Walter1988,Bouvier1986}.  As \citet{Clarke2000} have found evidence for differing stellar rotation distributions among T Tauri stars in Taurus and Orion, we have attempted to construct our sample exclusively from PMS in the same regions studied by D05.  This restricts the PMS sample to Tau-Aur and $\rho$ Oph since there are no well studied PMS stars in Ser and the D05 study includes only 5 objects in other regions, too few for statistical comparisons. We also restricted the comparison sample to stars of spectral type K0 and later to ensure that it sampled the same range of effective temperatures (T$_{eff}$) as the protostars in our sample (see Fig. 1).

\begin{figure*}
\plotone{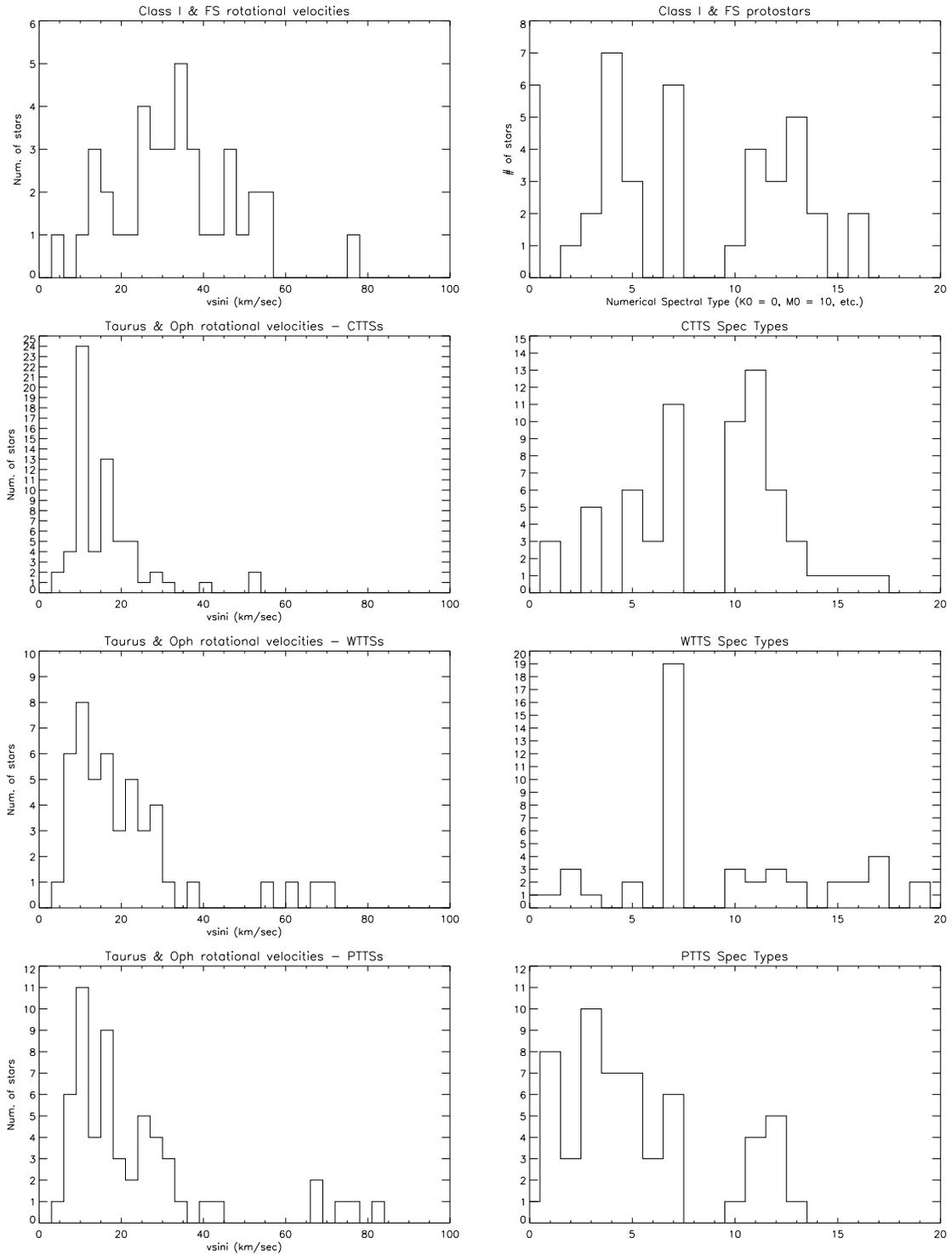}
\caption{\scriptsize{Display of the $v$sin$i$ and numerical spectral type distribution (K0 = 0, K5 = 5, M0 = 10, M5 = 15).  Each tier displays the sample for a given evolutionary phase, with the rotation velocity distribution for that phase on the left, and the distribution of spectral types for stars making up that sample shown on the right.}}
\label{fig2}
\end{figure*}

We have also divided these stars according to their pre-main sequence evolutionary phase; each star is classified as either a Classical T Tauri Star (CTTS), a Weak T Tauri Star (WTTS), or Post T Tauri Star (PTTS), with classifications largely adopted from the primary reference for an individual source. The notable exception concerns those objects detected as X-ray bright sources in the ROSAT All-Sky Survey \citep{Neuhaeuser1995,Wichmann2000}; given the uncertainty concerning the true ages of these objects \citep{Briceno1997,Briceno1999} we classify them as PTTSs by default unless they have been identified by other criteria (H$\alpha$ equivalent width, IR excess, etc.) as a CTTS or WTTS.  Projected rotation velocity distributions for these three comparison samples, along with the protostellar $v$ sin $i$ distribution shown in Figure 1, are shown in Figure 2.  Note that all samples (with the exception of PTTSs) contain objects with spectral types ranging from K0 to mid M, and the median spectral type for each sample (again excluding PTTSs) is at the late K/early M boundary.  While noting the accretion uncertainties raised in the previous section, we interpret the similar spectral type distributions of the various samples as a sign that each sample represents a relatively similar sampling in final stellar mass.

\section{Calculating Equatorial Rotation Velocity Distributions}

While analysis of spectroscopic lines can produce an estimate of $v$ sin $i$, without knowledge of the photometric period or other inclination diagnostics, it is impossible to solve the degeneracy between $v$ and $i$ for any particular star.  Assuming that the rotation axes of a sample of stars with measured $v$ sin $i$ values are randomly oriented allows one to directly calculate the mean equatorial rotation velocity, $\bar{v}$, from the mean of the observed values, $\bar{y}$, where $y = v$ sin $i$.  As shown first by \citet{Chandrasekhar1950}, the means of the two distributions can be related as:
\begin{equation}
\label{moments}
\bar{v} = \frac{4}{\pi}\bar{y}
\end{equation}
\citet{Chandrasekhar1950} also demonstrate that, in principle, the exact shape of the equatorial rotation velocity frequency function, $f(v)$ (which we will refer to as the `true' frequency function), can be derived by inverting the relationship between $f(v)$ and the observed rotation velocity frequency function, $\phi(y)$.  Accordingly, $f(v)$ can be written as:
\begin{equation}
\label{truefromobs}
f(v) = - \frac{2}{\pi} v^2 \frac{\partial}{\partial v} v \ \int_v^{\infty} \frac{\phi(y)}{y^2 \sqrt{y^2 - v^2}} dy \ 
\end{equation}
which can be numerically calculated, as shown by \citet{Gaige1993}, given a continuous form of $\phi(y)$.  The remainder of this section merely describes our implementation of the method originally presented by \citet{Gaige1993}: interested readers are encouraged to consult the primary work for a fuller explanation.

Our observations are simply the rotation velocities of a discrete sample of stars.  These data are inherently dis-continuous, even when formed into a histogram as in Figure 1.  To translate this set of N discrete observations into a continuous frequency function, $\phi(y)$, we sum a set of gaussians centered on each observed rotation velocity ($Y_i$) with a width set by the uncertainties in the derived velocities ($\sigma =$ 4 km s$^{-1}$), as shown in Equation 3:
\begin{equation}
\label{makecont}
\phi(y)=\frac{1}{N}\sum_{i=1}^{N}\frac{1}{\sigma\sqrt{2\pi}}(e^{\frac{-(y-Y_i)^2}{2\sigma^2}} + e^{\frac{-(y+Y_i)^2}{2\sigma^2}})
\end{equation}
With a continuous observed frequency function in hand, we are now able to apply the numerical inversion algorithm presented by \citet{Gaige1993} in their Equation 35 to solve our Equation 2 and calculate the `true' frequency function, $f(v)$.

To test that we have accurately implemented this numerical inversion technique, we have created simulated observed velocity distributions as inputs for this inversion method, and compared the results with the underlying `true' equatorial velocity distributions originally used in generating the simulated observed sample.  To create an simulated observed velocity distribution as input for the numerical inversion algorithm, we first randomly selected a sample of N stars from a given equatorial rotation velocity distribution.  Each star was then assigned a random inclination, selected in a manner consistent with an isotropic distribution of rotational axes, and the resultant simulated observed $v$ sin $i$ for that star is calculated.  All N artifically observed velocities were then combined into an aggregate velocity distribution; this distribution was analagous to the distribution of velocities from D05 which form the raw observational data for this study.  

To complete our test of the inversion method, we then completed the necessary steps to derive a `true' equatorial rotation velocity distribution from an observed set of data: a continuous frequency function was formed from the discrete set of simulated observed velocities, and this continuous frequency function was then numerically inverted following Equation 35 of \citet{Gaige1993}.  We tested the method with two types of input velocity distributions: a sample of 1000 stars with a single allowed equatorial velocity ($v_{eq} =$ 60 km s$^{-1}$) and a sample of 1000 stars where equatorial velocities were randomly distributed with a uniform probability within the range 0 km s$^{-1} < v <$ 60 km s$^{-1}$.  The basic form of the inverted 'true' distribution is correct in each case, deriving a single velocity population for the first case and a more broadly distributed population in the second case (shown in Figure 3).  All differences between input and output distributions were consistent with the gaussian smoothing and finite statistical noise of the input samples.  

\begin{figure}
\plotone{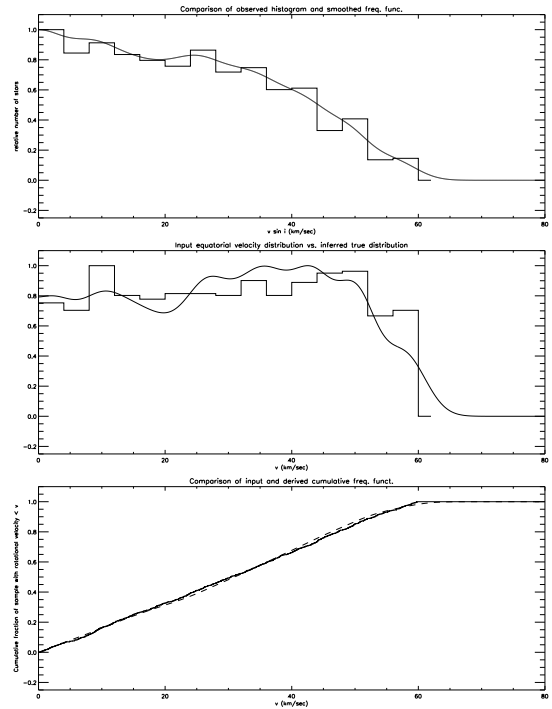} 
\caption {\scriptsize{A test of the accuracy of the numerical inversion technique used to determine the 'true' distribution of equatorial rotational velocities from a set of observed $v$ sin $i$ values.  These tests were run on a sample of 1000 synthetic stars with velocities distributed uniformly between 0 and 60 km s$^{-1}$ and random inclinations.  Top panel: A comparison between a histogram of the simulated 'observed' vsini values and the smoothed, continuous rotational frequency function, $\phi(y)$.  Middle panel: A comparison between a histogram of the input equatorial rotational velocities and the numerically inverted determination of the 'true' distribution function.  Bottom panel: a comparison of the cumulative frequency functions for both the input equatorial velocity distribution (solid) and the derived 'true' cumulative frequency function (dashed).}}
\label{fig3}
\end {figure}

The middle panel of Figure 3 shows the amplitude of statistical fluctuations in both the input equatorial rotation velocity distribution and the inverted `true' distribution.  The input distribution was selected such that all velocities had a uniform probability, so deviations from a constant number of objects per bin represent  statistical fluctuations in the sampling of the underlying distribution, consistent with the level expected by Poisson statistics.  This random velocity sampling, coupled with each star's randomly assigned inclination, results in some level of disagreement between the `observed' sample and an ideally sampled, perfectly isotropic distribution of rotational axes required for the method to return the correct result.  Inspection of the middle panel of Figure 3 indicates that, even for well sampled populations, deviations between the inverted `true' distribution and the probability function from which stars were selected may be as high as $\pm$15$\%$ at any individual velocity.  

The cumulative velocity distributions for both the input and inverted `true' distributions are shown in the bottom panel of Figure 3.  These distributions, which display the integral of the probability distribution from 0 to a given velocity, appear much less noisy to the eye.  Individual peaks and valleys from the middle panels are smoothed over by the integration, resulting in a good degree of agreement between the input and inverted `true' distributions, within the limits of the resolution of the technique.

Having tested the accuracy of the inversion routine in the well sampled regime, we also simulated the results obtained from sparsely sampled simulated observed distributions, to test the statistical fluctuations which could be introduced into our method as a result of using an input sample of only 38 stars.  This test shows the same types of errors as in the N$=$ 1000 test, but amplified, as statistical fluctuations in the input sample are correspondingly larger as a fraction of the number of stars within a given range of velocities.  The underlying form of the derived `true' distribution still appears fundamentally correct, as a broadly distributed population, but statistical fluctuations in the resultant probability distribution appear to have increased to $\pm$ 30$\%$ at any given velocity.  However, the cumulative frequency function continues to integrate over many of the sharp peaks and valleys in the inverted probability distributions, and the agreement between the two functions appears good, even with only N$=$38 objects to base the inversion on.  

We conclude from these tests that the underlying inversion method is sound and has been implemented correctly, though our sparsely sampled input distribution will clearly affect the accuracy of our derived `true' distribution.  Thus, in Figure 4 we present the numerical inversion of the observed $v$ sin $i$ distribution derived from the 41 protostars in our sample.  We expect from our previous tests that statistical fluctuations on the order of $\pm$30$\%$ \textit{at any given equatorial rotation velocity} will exist in our derived probability function (middle panel).  Our cumulative frequency function (lower panel), however, integrates over many of these fluctuations and presents a more statistically robust assesment of the distribution of equatorial rotational velocities for CI/FS protostars.  

\begin{figure}
\plotone{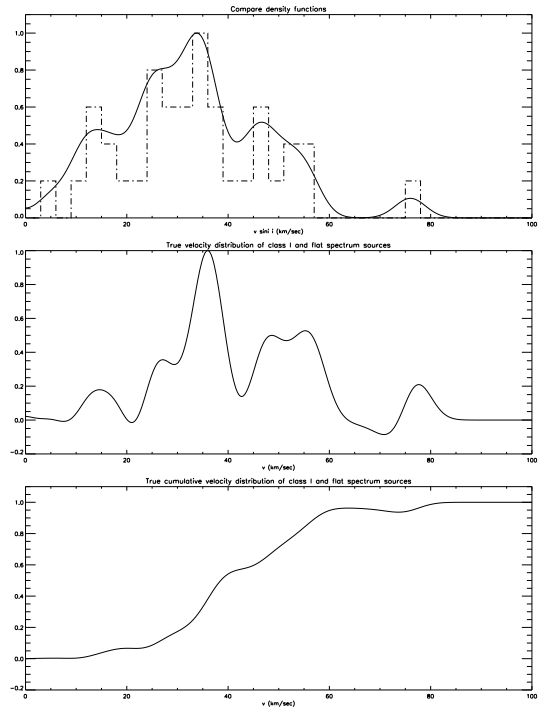} 
\caption {\scriptsize{The inversion of the $v$ sin $i$ distribution measured by D05.  Top panel: a comparison between the observed $v$ sin $i$ distribution (dashed histogram) and the resultant continuous frequency function, $\phi(y)$.  Middle panel: The inverted `true' frequency function, showing the probability of a class I or flat spectrum protostar having a given equatorial rotation velocity v.  Bottom panel: The cumulative frequency function of the `true' equatorial rotation velocity distribution of class I and flat spectrum protostars. }}
\label{fig4}
\end {figure}

For comparison, Figure 5 presents the results of this numerical inversion technique as applied to our sample of CTTS rotational velocities obtained from the literature. 

\begin{figure}
\plotone{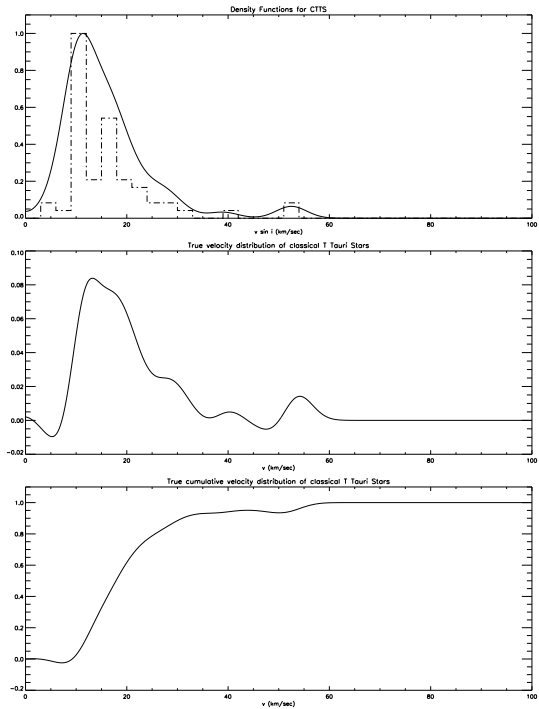} 
\caption {\scriptsize{An inversion of the observed $v$sin$i$ frequency function to produce the `true' frequency function, as in Figure 4, but for the CTTS comparison sample.}}
\label{fig5}
\end {figure}

\section{Analysis}
\subsection{Statistical tests of angular momentum evolution and possible sample biases}
A first test for the presence of angular momentum evolution between the CI/FS sample and that of more visible, and presumably more evolved, PMS stars can be determined from the observed $v$ sin $i$ distributions contained in Figure 2.  The samples of observed $v$ sin $i$s taken from the literature for each T Tauri stage can be compared with the observed $v$ sin $i$s of the CI/FS sample via a two sided Kolmogorov-Smirnov (KS) test for similarity.  The KS test compares the cumulative frequency functions of the observed rotation velocity distributions for the two samples; the maximum distance between the two cumulative frequency functions can be used to calculate the probability the distributions were drawn from the same parent sample \citep{Press1986}.  The KS test reports a $1.4 \times 10^{-10}$, $1.6 \times 10^{-5}$ and $1.6 \times 10^{-4}$  probability that the CTTS, WTTS and PTTS samples, respectively, were selected from the same parent distribution as the CI/FS sources.  

The KS tests simply confirm statistically what can be seen by comparing the distributions in Figure 2 by eye: the mean rotation velocity of the CI/FS sample is considerably higher than that of the PMS samples, and the shapes of the two distributions differ as well.  The PTTSs are the most similar of the PMS samples to the CI/FS sample, a fact attributable to the higher rotation rates reached by the PTTSs as they contract onto the main sequence with little angular momentum loss.  To investigate the amount of angular momentum evolution which must have taken place between the CI/FS and the CTTS phases, we select as representative velocities of each phase the velocity at which the `true' cumulative velocity distribution reaches the 50$\%$ level.  These velocities are $\sim$ 38 km s$^{-1}$ and $\sim$ 18 km s$^{-1}$, respectively.  

To seek evidence for the nature of any protostellar angular momentum regulation mechanism at work during the CI/FS phase, we have searched for correlations between projected rotation velocity and other observable protostellar parameters indicative of their evolutionary state.  In doing so, we have found no dependence of projected rotation velocity on accretion rate as measured by either Br $\gamma$ emission or excess continuum emission (r$_k$; see Figure 23 by D05).  There also appears to be no dependence of rotation velocity on the slope of the SED ($\alpha$) between 1 and 10 $\mu$m for Class I or flat spectrum sources.

The only correlation which does appear to be present is one between projected rotation velocity and T$_{eff}$, as shown in Figure 6.  The correlation has a large amount of scatter, represented by the fairly low (0.53) correlation coefficient.  Removing the two high temperature, high $v$ sin $i$ objects further weakens the correlation (correlation coefficient = 0.33), though the slope of the best fit line appears relatively similar between the two fits.  

\begin{figure}
\plotone{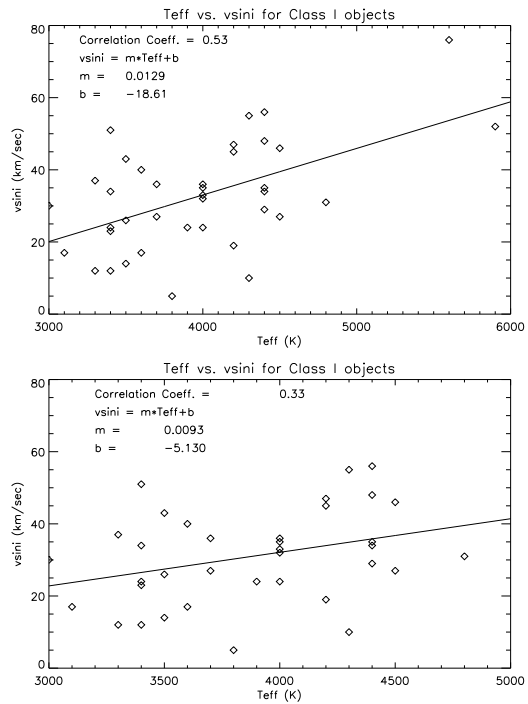} 
\caption {\scriptsize{Top: T$_{eff}$ vs. vsini for Class I/flat spectrum sources are plotted as diamonds for individual sources.  The best fit linear least squares relationship is overplotted as a line; the raw correlation coefficient of the data, as well as the coefficients of the best fit line are displayed within the plot.  Bottom: As above, but after removing the two high temperature, high vsini objects from the sample.}}
\label{fig6}
\end {figure}

It is important to note that both the CI/FS and CTTS samples suffer from observational biases which could skew the resultant $v$ sin $i$ distributions.  In the case of the CI/FS sample, binarity (tested for and previously described in section 2.1) and veiling biases are likely present at some level within the sample.  A veiling bias will result from difficulty in detecting photospheric absorption lines from CI/FS sources whose spectra is affected by strong rotational broadening or excess continuum emission at 2$\mu$m from circumstellar material.  Both the broadening due to rapid rotation and the veiling caused by excess continuum emission make photospheric lines appear shallow, and thus harder to detect in a spectra with a finite signal-to-noise ratio.  Therefore, the fact that 11 of the 52 Class I/flat spectrum sources targeted for the survey of D05 did not reveal detectable photospheric absorption lines indicates that those sources were either a) rapid rotators, b) heavily accreting sources, or c) earlier than G type, where the photospheric absorption features analyzed here become too weak to detect.  The last possibility, that these stars are merely hotter than those which show photospheric absorption lines at 2 $\mu$m, can be ruled out for most of the sample on the basis of a lack of a strong HI Br $\gamma$ absorption line, a common spectral feature of hot stars.  None of the stars for which we fail to detect photospheric absorption features possessed a Br $\gamma$ absorption line, and 9 of the 11 had Br $\gamma$ emission features in excess of 1 $\AA$ equivalent width, widely interpreted as a sign of active accretion.  As a result, the detections of D05 are incomplete at the 20$\%$ level and likely biased against the most heavily accreting or, most significantly for this work, rapidly rotating protostars; see Figure 23 of D05 for the region of r$_{K}$ and $v$ sin $i$ parameter space to which the survey is sensitive.  Our characteristic velocity for this phase is thus likely an underestimate of the median velocity of a truly complete sample of CI/FS sources in the absence of binary contamination.

The CTTS sample is biased in the opposite sense, such that our estimate of the characteristic velocity of this phase is an overestimate.  This bias results from the observational limits of the surveys used to measure many of the stars within the sample, which were only sensitive to rotational velocities $>$ 10 km s$^{-1}$.  Thus, the large peak in the CTTS distribution at 10 km s$^{-1}$ is formed with a number of stars whose $v$ sin $i$ detection is merely an upper limit; replacing those velocities with the true, and presumably lower, rotation velocity would lower the characteristic velocity we have determined for the CTTS phase.  

As the CTTS sample is biased towards rapid rotators, and the Class I/flat spectrum sample is biased towards slower rotators, the $\sim$ 20 km s$^{-1}$ difference between the characteristic equatorial velocities of each phase represents a lower limit to the true change in equatorial rotation velocity evolution between the two phases.  

\subsection{Comparison with previous work}

In a recent study of optical spectra of 11 Class I sources and 21 Class II sources (all known to be driving Herbig Haro outflows) in Taurus, WH04 found no evidence for differences in the rotation properties of the two samples.  As IR selection of Class II objects and optical selection of CTTSs are thought to identify identical populations of pre-main sequence stars, WH04 appear to draw an opposite conclusion to that found in \S 4.1.  Deriving rotation velocities from analysis of optical spectra showing photospheric absorption features, WH04 find nearly identical means for the two samples.  The mean of the Class I sample was $<v$ sin $i>$ = 18.95 km s$^{-1}$ while the mean of the Class II sample was $<v$ sin $i>$ = 17.39 km s$^{-1}$ such that a KS test revealed that there is 7.9$\%$ of both samples sharing the same parent distribution.  The Class II sample does return a lower median $v$ sin $i$ (14.4 km s$^{-1}$) than the Class I sample (18 km s$^{-1}$).  As shown previously in \S 2.1, $v$ sin $i$ determinations for the 5 stars in common between this program and that of WH04 agree well: 4 in 5 have determinations identical within the errors and another object (IRAS 04106+2610) is only marginally detected (S/N = 2.1) by WH04 with a $v$ sin $i$ value flagged as being highly uncertain.  Thus, both samples appear to agree well on $v$ sin $i$ for common objects with observations in the high S/N regime.  

While measurements of $v$ sin $i$ for individual sources do appear to match well for those objects observed by both programs, the mean projected rotation velocity for Class I objects in Taurus derived by each program does differ (WH04: 19.1 km s$^{-1}$, this work: 30.1 km s$^{-1}$).  This difference appears to be due to an inverse correlation between the rotation rate of the Class I object and its optical visibility.  Three Taurus Class I objects for which we derive high rotation rates from well detected infrared absorption features (IRAS 04016+2610 [46 km s$^{-1}$], IRAS 04295+2251 [51 km s$^{-1}$] and L1551+IRS5 [31 km s$^{-1}$]) appear to be so faint in the optical that WH04 were unable to derive confident estimates of $v$ sin $i$.  Accordingly, we conclude that the difference between the Class I $v$ sin $i$ distribution presented here and that of WH04 can be attributed to the fact that the fastest rotating Class I sources appear to be optically dim, likely due to higher levels of extinction, consistent with our previous finding that more heavily embedded Class I sources have higher mean $v$ sin $i$ values than more optically visible Class II/CTTS sources\footnote{We note that protostellar envelopes may be flattened along their rotation axes, either due to rotation itself or from clearings produced by bipolar jets.  If this is the case, extinction will be inclination dependent, and a bias towards optically visible sources will result in a bias towards sources with small sin $i$s, and thus small $v$ sin $i$s.}.  

\subsection{Rotation velocity comparisons between regions}

The bulk of the information currently known concerning the rotation velocities of CTTSs has resulted primarily from studies of two star formation regions: Taurus \citep{Bouvier1986,Hartmann1989,Bouvier1990} and Orion \citep{Stassun1999,Rebull2001,Rhode2001,Sicilia-Aguilar2005}.  Recent work by \citet{Clarke2000} suggests that rotation properties for samples of stars selected as CTTSs in each region may be significantly different, while \citet{Sicilia-Aguilar2005} find a typical CTTS rotation velocity in Orion which agrees well with our comparison sample of CTTSs drawn from Taurus and $\rho$ Ophiuchi, implying a relatively small role for environmental effects on the angular momentum evolution of young stars.  With measurements of rotation velocity for $\sim$ 10 Class I protostars in each of three star formation regions ($\rho$ Ophiuchi, Serpens and Taurus), it is possible to begin to probe the degree to which the rotational properties of Class I sources differ across regions.  

Of the three regions, Serpens has the largest mean observed $v$ sin $i$ in our sample (36.8 km s$^{-1}$), followed by $\rho$ Ophiuchi (31.1 km s$^{-1}$) and finally Taurus (30.1 km s$^{-1}$).  Two sided KS tests indicate Taurus and Serpens differ most strongly, but still have a 18\% chance of sharing a parent distribution, so these differences between rotational properties of Class I objects in differing star formation regions are not yet statistically significant.  

\section{Angular Momentum Discussion}  
The total angular momentum (L$_{tot}$) of a star can be expressed as $L_{tot} = I_{*} \omega $, where I$_*$ is the stellar moment of inertia and $\omega$ is the angular velocity of the star.  Assuming spherically symmetric, solid body rotation of a protostar with a convective interior, I$_* \sim 0.2 M_* R_*^2$ \citep{Claret1990}, and $\omega = \frac{v}{R_*}$.  Thus, the angular momentum content of a given star can be simply written as L$_{tot} \sim 0.2 M_* R_*v$. This functional form provides a simple physical framework in which to interpret the implications of the difference in the median equatorial velocities for both the CTTS and the CI/FS samples under various theoretical frameworks for protostellar evolution.

\subsection{Evolution With No Extraction of Protostellar Angular Momentum}  For there to be no extraction of angular momentum from the protostar between the CI/FS and CTTS phases, the difference of a factor of two in the characteristic median equatorial velocities requires that the product of M$_*$R$_*$ must be double that of the CI/FS stage by the CTTS phase.  Absent any significant changes in the internal structure of protostars between the two phases (such as the onset of dueterium shell burning, expected to inflate the stellar radius by a factor of two or greater but only occur once stars have grown to M$_* > $ 3 M$_{\odot}$; Palla 1999), R$_*$ grows slowly as the total mass of the star increases: R$_* \propto$ M$_*^{\frac{1}{3}}$.  Stellar radius is believed to \textit{decrease}, not increase, as PMS stars evolve toward the main sequence, but this is not a significant effect in the evolution from the CI/FS phases; D05 find that the CI/FS sample had surface gravities similar to those of CTTSs.  Thus, an increase in M$_*$R$_*$ by a factor of two and no significant changes in internal structure implies an increase in M$_*$ by a factor of $\sim$ 1.7.  Assuming a characteristic mass of 1 M$_{\odot}$ for our sample, as predicted by the models of \citet{Baraffe1998} for a T$_{eff} =$ 3800 K star at 1 Myr, an increase of this size implies accretion of 0.7 M$_{\odot}$ of material.

CI/FS objects are commonly thought to be nearing the end of the main mass accretion phase (a phase fully completed by the Class II/CTTS stage), which should rule out a near doubling of the protostellar mass.  This conclusion is butressed by the comparison of spectral types for the CI/FS and CTTS samples (as seen in Figure 2 and discussed in \S 2), consistent with little to no significant additional mass accretion between the two phases.  However, even if mass accretion takes place at a significant level between the CI/FS phase and the CTTS phase, allowing that additional mass to lead to a decrease in rotation velocity requires that the newly accreted material have significantly lower specific angular momentum than that which was previously accreted by the protostar.  \citet{Popham1996} suggest that FU Ori-like accretion events can result in mass accretion distinguished by markedly lower specific angular momentum than that which is accreted during more quiescent periods of protostellar evolution, precisely the effect needed to drive the rotation velocity of CTTSs down from the CI/FS level.  However, the amount of FU Ori accretion required would be unusual given current estimates of the FU Ori duty cycle: phases of rapid mass accretion at the 10$^{-4}$ M$_{\odot}$ yr$^{-1}$ level \citep{Hartmann1996} would require a cumulative total of $\sim$10$^4$ years to be spent in the high accretion state to increase the protostellar mass by $\sim$1 M$_{\odot}$, a factor of 20 greater than the current estimated FU Ori duty cycle of $\sim$500 years in the high accretion state for each low mass star.

Ultimately, a model where the change in equatorial rotation velocity observed between the Class I and CTTS phases is due solely to redistribution of the current protostellar angular momentum is difficult to construct.  This model is also difficult to reconcile simultaneously with current measurements of both Class I and FU Ori accretion rates and lifetimes, casting doubt on the assumption that no extraction of angular momentum takes place during the Class I phase.  

\subsection{Deriving \.J assuming efficient extraction of angular momentum}

The opposite limiting case to that presented in \S 5.1 would hold that the difference in $v$ sin $i$s of the two samples is due completely to extraction of angular momentum from the protostar.  We thus investigate the amount and rate of angular momentum extraction implied under a model where the physical parameters (M$_*$, R$_*$) of a typical CI/FS object identically match those of a typical CTTS.  Assuming canonical values of M$_* = 1.0$ M$_{\odot}$ and R$_* = 2.7$ R$_{\odot}$ and expressing the change in angular momentum between the two phases as $\Delta $J$ = 0.2M_*R_*\Delta v$, we find $\Delta J = 1.5 \times 10^{50}$ gm cm$^2$ s$^{-1}$.  Further assuming a mean age of $10^5$ years for our CI/FS sample and $10^6$ years for our CTTS sample implies $\sim$ 0.9 Myrs of elapsed time between the Class I and CTTS phases, and allows a calculation of \.J $= \frac{\Delta J}{\Delta t} = 5.3 \times 10^{36}$ gm cm$^2$ s$^{-2}$.  This derived value for \.J assumes no particular model for how angular momentum extraction is performed, but depends only on a) the assumed separation in ages between the Class I and CTTS phases [.9 Myrs], b) the assumed values for M$_*$ and R$_*$ [M$_*$ = 1.0 M$_{\odot}$, R$_*$ = 2.7 R$_{\odot}$], as well as assuming those values characterize both the CTTS and CI/FS phases, and c) the derived difference [$\sim$ 20 km s$^{-1}$] between typical CI/FS and CTTS equatorial rotational velocites.
 
\subsection{Extraction of angular momentum via disk locking}
Disk locking models describe extraction of protostellar angular momentum via magnetic coupling of the protostar through accretion columns to a region in the disk whose Keplerian angular velocity is slightly lower than the angular velocity of stellar rotation.  Drag on the stellar magnetic field lines as they rotate through the disk then begins to slow the stellar rotation until the star, and likewise its magnetic field, are rotating with the same angular velocity as the disk.  Once the protostar and coupling radius of the disk are rotating with the same angular velocity, the magnetic fields are stationary in the rotating reference frame of the disk -- no magnetic drag occurs and no additional angular momentum is extracted from the protostar \citep{Shu1994,Konigl1991}.  Observational evidence of excess blue continuum emission (assumed to arise from accretion shocks; Bertout, Basri \& Bouvier 1988) and the presence of magnetic fields strong enough to couple protostars to the ionized portions of their disks \citep{Johns-Krull1999} have been interpreted in support of this paradigm.  However, recent models \citep{Matt2004} which attempt to account for dynamical evolution of magnetic field topologies in response to the star disk interaction appear to have difficulty extracting angular momentum at the rates assumed necessary to account for the slow rotation rates ($\sim$ 10\% of breakup) observed in CTTSs.  Additionally, observational evidence for the correlations predicted by these theories is mixed \citep{Johns-Krull2002}.  Despite these difficulties, disk locking theories present the most fully developed theoretical framework within which to explore the implications of the rotation velocity evolution of protostars as they pass from the CI/FS into the CTTS phase.  

\subsubsection{The typical Class I co-rotation radius}

We determine here the corotation radius of the protostellar disk to which a typical CI/FS object would be magnetically coupled if fully disk locked.  We note that the corotation radius will be a lower limit to the radius at which the protostar couples to the disk, as the time scale over which disk locking removes angular momentum from a protostar may be sufficiently long that these protostars are still being magnetically braked \citep{Hartmann2002}, requiring coupling radii outside the co-rotation radius.  It should be noted, however, that differential rotation between the star and disk should drive significant changes in the magnetic field topology, leading to reconfigurations that lessen the efficiency of angular momentum transport \citep{Matt2004}. 

To derive the typical co-rotation radius implied for the 38 km s$^{-1}$ median velocity of CI/FS sources, we simply invoke Kepler's third law, assuming circular disk orbits where the mass of the disk has a negligable effect on the gravitational potential of the system, and thus $P_{disk}^2 = \frac{4 \pi^2}{G M_*} r^3$.  As we seek the radius of the disk where the Keplerian period is equivalent to the stellar rotation period, we can thus express the corotation radius as $r_{co} = (\frac{R_*^2 G M_*}{v^2})^{1/3}$.  Assuming a canonical mass of 0.5 M$_{\odot}$ and radius of 3.1 R$_{\odot}$ for our CI/FS sample (we adopt 0.5 M$_{\odot}$ and 3.1 R$_{\odot}$ here instead of the 1.0 M$_{\odot}$ and 2.7 R$_{\odot}$ assumed in previous sections to facilitate comparison with the results of \citet{Greene2002}; assuming a mass of 1.0 M$_{\odot}$ and radius of 2.7 R$_{\odot}$ here would lead to an increase in the derived radius by 15\%) implies a co-rotation radius of 8.6 R$_{\odot}$, or 2.8 R$_*$.  We note that this value is close to the 2.1 R$_*$ magnetic coupling radius derived for the Class I object YLW 15 A by \citet{Greene2002}.  The \citet{Greene2002} result was determined by simultaneously calculating the mass accretion rate and magnetic coupling radius necessary to account for the observed accretion luminosity using the magnetic accretion models of \citet{Shu1994} with protostellar parameters identical to those presented above.  The agreement between these parameters suggest that the location at which co-rotation occurs in a typical CI/FS object matches well with the location at which coupling would lead to the observed rotation rate of CI/FS objects if magnetic torques are strong enough to lock the protostellar disk to the protostar.  

\subsubsection{Implied protostellar mass accretion rates}

Magnetospheric accretion models typically assume that the circumstellar disk is truncated at the Alfv\'en radius, where the energy density of the stellar magnetic field is matched by the kinetic energy density of the infalling material.  This truncation radius (r$_t$) can be expressed as

\begin{equation}
\label{alfven}
r_t \simeq R_* (\frac{B_*^4 R_*^5}{G M \dot{M}^2})^{\frac{1}{7}}
\end{equation}

 It is also near this truncation radius that magnetospheric accretion models typically locate the co-rotation point to which the star is locked, such that $\omega_{*} = \omega_{disk}$.  Thus, for a star which is fully disk locked to the Keplerian orbital velocity at the disk truncation point, we can express the stellar rotation velocity as 
\begin{equation}
\label{disk speed}
v_* = v_{disk}(r_t) \frac{R_*}{r_t} = \sqrt{\frac{G M R_*^2}{r_t^3}}
\end{equation}

Combining equations 4 and 5 allow us to express v$_*$ in terms of fundamental protostellar parameters, such that 
\begin{equation}
\label{star v}
v_* \simeq (\frac{G^5 M_*^5 \dot{M}^3 }{R_*^{11} B_*^6})^{\frac{1}{7}}
\end{equation}

Therefore, viewed in the context of magnetospheric accretion, along with the assumption that the basic stellar parameters (M$_*$, R$_*$, B$_*$) do not change significantly between the CI/FS and CTTS phases, the observed change in the mean rotation velocity of the two samples would naturally be attributed to a change in the mean protostellar accretion rates between the two phases if the protostars are fully disk locked.  As $v_* \propto $ \.M$^{\frac{3}{7}}$, a decrease by a factor of 2 in the characteristic rotational velocities from the CI/FS to CTTS phase would represent a factor of 5 decline in the typical accretion rate from the CI/FS to the CTTS phase.  This finding fits well into the observational paradigm which assumes CI/FS objects are younger, more heavily accreting protostars than CTTSs.  We must note, however, that the stellar rotation velocity is more weakly dependent on \.M than any other stellar parameter: thus, the degree to which stellar parameters change between the CI/FS and CTTS phases must be investigated before protostellar rotation velocity can serve to test the validity of magnetospheric accretion models by comparing rotation velocities to observed accretion rates.   

As noted by \citet{Hartmann2002}, models which assume a transport of angular momentum away from a protostar through a viscous circumstellar disk require inward accretion of mass in the disk to allow the outward transport of angular momentum.  We can thus attempt a crude estimation of the typical mass accretion rate between the CI/FS and CTTS phases by estimating the mass accretion rate necessary to allow the \.J derived in \S 5.2.  We equate the rate of change of the angular momentum of the protostar with the rate at which angular momentum is transported away from the co-rotation radius via accretion through the disk.  Following Hartmann's equation 5, but ignoring the spin-up torque of the additional newly accreted material to derive a lower limit to the needed accretion rate, we find: 

\begin{equation}
\label{jdot}
\dot{J} \approx \dot{M} \omega(r_{co}) r_{co}^2 = \dot{M} \frac{v_*}{R_*} r_{co}^2
\end{equation}

Applying a corotation radius of 9.9 R$_{\odot}$ derived in \S 5.3.1 for a 1 M$_{\odot}$ protostar with a 2.7 R$_{\odot}$ stellar radius (R$_*$) and a 28 km s$^{-1}$ stellar velocity chosen to be intermediate to the 38 km s$^{-1}$ mean CI/FS velocity and the 18 km s$^{-1}$ mean CTTS velocity, we can solve for the $\dot{M}$ applicable to objects transitioning from the CI/FS phase to CTTSs.  We note that this calculation implicitly assumes that \.M decreases slowly enough that the star stays magnetically coupled to its disk over the CI/FS to CTTS transition period, allowing continuous transfer of angular momentum.  We therefore find a lower limit to the mass accretion rate between the CI/FS and CTTS phases of \.M $\sim 10^{-8}$ M$_{\odot}$ yr$^{-1}$ would be required for a viscous disk to transport the angular momentum our observations require to be extracted between the two protostellar phases under the assumptions of \S 5.2.  This lower limit fits well with the accretion rates measured for CTTSs \citep[$ \sim 10^{-8}$ M$_{\odot}$ yr$^{-1}$;][]{Muzerolle1998}.  This accretion rate lower limit would increase if the age gap between the CI/FS and CTTS phases were lowered from the assumed value of 0.9 Myrs, or if the effects of the spin-up torque due to accreted material were accounted for; either of these effects would drive the intermediate accretion rate closer to the 10$^{-6}$ M$_{\odot}$ yr$^{-1}$ estimated for the protostar YLW 15 A in the context of magnetospheric accretion models \citep{Greene2002}.

Lastly, \citet{Hartmann2002} also notes that the timescale over which the disk-locking process removes angular momentum from a young star is a dynamic quantity, sensitive to changes in the star's rotation velocity, magnetic field strength, mass, and mass accretion rate.   Specifically, \citet{Hartmann2002} predicts in his equation 8 a disk-locking timescale which varies as:
\begin{equation}
\label{timescale}
\tau_{DB} \gtrsim 4.5 \times 10^6 yr M_{0.5} \dot{M_{-8}^{-1}} f,
\end{equation}
where $M_{0.5}$ is the stellar mass in units of half a solar mass, $\dot{M_{-8}}$ is the mass accretion rate in units of $10^{-8} M_{\odot} yr^{-1}$ and f is the stellar rotation velocity expressed as a fraction of the star's breakup velocity ($\sim$ .15 for the CI/FS objects reported here).  The large mass accretion rates believed to exist during the CI/FS stage ($\dot{M_{\odot}} \sim 10^{-6} M_{\odot} yr^{-1}$; Greene \& Lada 2002) thus predicts a disk-locking timescale on the order of $10^4$ years, an angular momemtum evolution timescale much shorter than that assumed in \S 5.2.  This timescale is sufficient to quickly remove significant amounts of angular momentum from the protostar and also consistent with the decrease in stellar rotation observed between the CI/FS and CTTS phases.  The reduced mass accretion rates characteristic of the CTTS phase, however, suggest a corresponding increase in the disk-locking timescale and thus less efficient extraction of angular momentum, consistent with the spin-up of the rotation rate observed by \citet{Sicilia-Aguilar2005} between the CTTS and WTTS phases.  A decline in the mass accretion rate of a protostar as it evolves through the Class I/CTTS/WTTS phases and the inverse dependence of the timescale of the disk locking phenomena on the mass accretion rate appear to mesh well with current observations of pre-main sequence rotation.

\section{Conclusions}

We have analyzed the projected rotation velocities ($v$ sin $i$s) of 38 D05 Class I/flat spectrum sources in nearby star formation regions with no apparent close companions, allowing a robust investigation of the angular momentum content of heavily embedded protostars.  The key results of this study are:  

\begin{enumerate}

\item{Class I/flat spectrum objects rotate more quickly (median equatorial rotational velocities $\sim$ 38 km s$^{-1}$) than CTTSs (median equatorial rotational velocities $\sim$ 18 km s$^{-1}$).  Viewed in context of the protostellar evolutionary paradigm whereby Class I/flat spectrum objects are the less evolved predecessors of the older CTTSs, it is hard to explain the change in rotation rates between the two phases without the presence of some mechanism which transfers angular momentum out of the protostar between the Class I and CTTS phases.}

\item{Projected rotation velocity ($v$ sin $i$) seems to be weakly correlated with T$_{eff}$ in our sample.  Projected rotation velocity does not seem to correlate with Br $\gamma$ emission (a common accretion tracer), the amount of excess continuum veiling (r$_k$), or the slope of the SED ($\alpha$) between the near and mid IR ($\alpha$).}

\item{Assuming Class I/flat spectrum sources possess physical characteristics (M$_*$,R$_*$,B$_*$) typical of pre-main sequence stars, fully disk locked Class I objects should have co-rotation radii within their protostellar disks that match well (within 30\%) with the predicted magnetic coupling radii of \citet{Shu1994} for similar sources.}

\item{The factor of two decrease in rotation rates between Class I/flat spectrum and CTTS sources, when interpreted in the context of disk locking models, implies a factor of 5 or greater drop in mass accretion rates between the two phases.}

\item{A lower limit of \.M $\sim 10^{-8}$ M$_{\odot}$ yr$^{-1}$ is required for objects transitioning from Class I/flat spectrum objects to CTTSs to account for the difference in rotation rates of the two classes by angular momentum extraction through a viscous disk and no changes in protostellar physical structure.  This lower limit implies a rate properly intermediate to typical estimates of Class I accretion rates (10$^{-6}$ M$_{\odot}$ yr$^{-1}$) and CTTS accretion rates (10$^{-8}$ M$_{\odot}$ yr$^{-1}$).}

\end{enumerate}

\acknowledgements

The authors wish to recognize and acknowledge the very significant cultural role and reverence that the summit of Mauna Kea has always had within the indigenous Hawaiian community.  We are most fortunate to have the opportunity to conduct observations from this mountain.  All data have been reduced using IRAF; IRAF is distributed by the National Optical Astronomy Observatories, which are operated by the Association of Universities for Research in Astronomy, Inc., under cooperative agreement with the National Science Foundation.  This research has made use of NASA's Astrophysics Data System Bibliographic Services, the SIMBAD database, operated at CDS, Strasbourg, France, and the VizieR database of astronomical catalogues \citep{Ochsenbein2000}.  K.R.C gratefully acknowledges the support of NASA grant 80-0273. All authors acknowledge support from NASA Origins of Solar Systems program via UPN 344-39-00-09 while conducting this work.  The authors thank the anonomyous referee for a quick, thorough and illuminating report whose comments have improved the work presented here.  

\clearpage

%--------------------------BIBLIOGRAPHY---------------------------

\clearpage

% This is where the figures go
% in the text refer to by label
% e.g. In Figure~\ref{fig-targetflags} we plot ...


\begin{thebibliography}{80}
\expandafter\ifx\csname natexlab\endcsname\relax\def\natexlab#1{#1}\fi

\bibitem[{{Baraffe} {et~al.}(1998){Baraffe}, {Chabrier}, {Allard}, \&
  {Hauschildt}}]{Baraffe1998}
{Baraffe}, I., {Chabrier}, G., {Allard}, F., \& {Hauschildt}, P.~H. 1998, \aap,
  337, 403

\bibitem[{{Barnes} {et~al.}(1999){Barnes}, {Sofia}, {Prosser}, \&
  {Stauffer}}]{Barnes1999}
{Barnes}, S.~A., {Sofia}, S., {Prosser}, C.~F., \& {Stauffer}, J.~R. 1999,
  \apj, 516, 263

\bibitem[{{Bodenheimer}(1995)}]{Bodenheimer1995}
{Bodenheimer}, P. 1995, \araa, 33, 199

\bibitem[{{Bontemps} {et~al.}(2001){Bontemps}, {Andr{\' e}}, {Kaas}, {Nordh},
  {Olofsson}, {Huldtgren}, {Abergel}, {Blommaert}, {Boulanger}, {Burgdorf},
  {Cesarsky}, {Cesarsky}, {Copet}, {Davies}, {Falgarone}, {Lagache},
  {Montmerle}, {P{\' e}rault}, {Persi}, {Prusti}, {Puget}, \&
  {Sibille}}]{Bontemps2001}
{Bontemps}, S., {Andr{\' e}}, P., {Kaas}, A.~A., {Nordh}, L., {Olofsson}, G.,
  {Huldtgren}, M., {Abergel}, A., {Blommaert}, J., {Boulanger}, F., {Burgdorf},
  M., {Cesarsky}, C.~J., {Cesarsky}, D., {Copet}, E., {Davies}, J.,
  {Falgarone}, E., {Lagache}, G., {Montmerle}, T., {P{\' e}rault}, M., {Persi},
  P., {Prusti}, T., {Puget}, J.~L., \& {Sibille}, F. 2001, \aap, 372, 173

\bibitem[{{Bouvier}(1990)}]{Bouvier1990}
{Bouvier}, J. 1990, \aj, 99, 946

\bibitem[{{Bouvier} {et~al.}(1986){Bouvier}, {Bertout}, {Benz}, \&
  {Mayor}}]{Bouvier1986}
{Bouvier}, J., {Bertout}, C., {Benz}, W., \& {Mayor}, M. 1986, \aap, 165, 110

\bibitem[{{Brice{\~ n}o} {et~al.}(1999){Brice{\~ n}o}, {Calvet}, {Kenyon}, \&
  {Hartmann}}]{Briceno1999}
{Brice{\~ n}o}, C., {Calvet}, N., {Kenyon}, S., \& {Hartmann}, L. 1999, \aj,
  118, 1354

\bibitem[{{Briceno} {et~al.}(1997){Briceno}, {Hartmann}, {Stauffer}, {Gagne},
  {Stern}, \& {Caillault}}]{Briceno1997}
{Briceno}, C., {Hartmann}, L.~W., {Stauffer}, J.~R., {Gagne}, M., {Stern},
  R.~A., \& {Caillault}, J. 1997, \aj, 113, 740

\bibitem[{{Chandrasekhar} \& {M{\" u}nch}(1950)}]{Chandrasekhar1950}
{Chandrasekhar}, S. \& {M{\" u}nch}, G. 1950, \apj, 111, 142

\bibitem[{{Claret} \& {Gimenez}(1990)}]{Claret1990}
{Claret}, A. \& {Gimenez}, A. 1990, \apss, 169, 215

\bibitem[{{Clarke} \& {Bouvier}(2000)}]{Clarke2000}
{Clarke}, C.~J. \& {Bouvier}, J. 2000, \mnras, 319, 457

\bibitem[{{Covey} {et~al.}(2005){Covey}, {Greene}, {Doppmann}, \&
  {Lada}}]{Covey2005b}
{Covey}, K.~R., {Greene}, T., {Doppmann}, G., \& {Lada}, C. 2005, AJ, 'in
  prep.'

\bibitem[{{D'Antona} \& {Mazzitelli}(1994)}]{D'Antona1994}
{D'Antona}, F. \& {Mazzitelli}, I. 1994, \apjs, 90, 467

\bibitem[{{de Jager} \& {Nieuwenhuijzen}(1987)}]{deJager1987}
{de Jager}, C. \& {Nieuwenhuijzen}, H. 1987, \aap, 177, 217

\bibitem[{{Doppmann} {et~al.}(2005){Doppmann}, {Greene}, {Lada}, \&
  {Covey}}]{Doppmann2005}
{Doppmann}, G.~W., {Greene}, T.~P., {Lada}, C., \& {Covey}, K. 2005, \aj,
  {submitted}

\bibitem[{{Doppmann} {et~al.}(2003){Doppmann}, {Jaffe}, \&
  {White}}]{Doppmann2003B}
{Doppmann}, G.~W., {Jaffe}, D.~T., \& {White}, R.~J. 2003, \aj, 126, 3043

\bibitem[{{Durisen} {et~al.}(2003){Durisen}, {Mejia}, \&
  {Pickett}}]{Durisen2003}
{Durisen}, R.~H., {Mejia}, A.~C., \& {Pickett}, B.~K. 2003, Recent Research
  Development in Applied Physics, 1, 173

\bibitem[{{Edwards} {et~al.}(1993){Edwards}, {Strom}, {Hartigan}, {Strom},
  {Hillenbrand}, {Herbst}, {Attridge}, {Merrill}, {Probst}, \&
  {Gatley}}]{Edwards1993}
{Edwards}, S., {Strom}, S.~E., {Hartigan}, P., {Strom}, K.~M., {Hillenbrand},
  L.~A., {Herbst}, W., {Attridge}, J., {Merrill}, K.~M., {Probst}, R., \&
  {Gatley}, I. 1993, \aj, 106, 372

\bibitem[{{Gaig\'e}(1993)}]{Gaige1993}
{Gaig\'e}, Y. 1993, \aap, 269, 267

\bibitem[{{Gameiro} \& {Lago}(1993)}]{Gameiro1993}
{Gameiro}, J.~F. \& {Lago}, M.~T.~V.~T. 1993, \mnras, 265, 359

\bibitem[{{Gray}(1992)}]{Gray1992}
{Gray}, D.~F. 1992, {The observation and analysis of stellar photospheres}
  (Cambridge Astrophysics Series, Cambridge: Cambridge University Press, 1992,
  2nd ed., ISBN 0521403200.)

\bibitem[{{Greene} \& {Lada}(1997)}]{Greene1997}
{Greene}, T.~P. \& {Lada}, C.~J. 1997, \aj, 114, 2157

\bibitem[{{Greene} \& {Lada}(2000)}]{Greene2000}
---. 2000, \aj, 120, 430

\bibitem[{{Greene} \& {Lada}(2002)}]{Greene2002}
---. 2002, \aj, 124, 2185

\bibitem[{{Greene} {et~al.}(1994){Greene}, {Wilking}, {Andre}, {Young}, \&
  {Lada}}]{Greene1994}
{Greene}, T.~P., {Wilking}, B.~A., {Andre}, P., {Young}, E.~T., \& {Lada},
  C.~J. 1994, \apj, 434, 614

\bibitem[{{Haisch} {et~al.}(2004){Haisch}, {Greene}, {Barsony}, \&
  {Stahler}}]{Haisch2004}
{Haisch}, K.~E., {Greene}, T.~P., {Barsony}, M., \& {Stahler}, S.~W. 2004, \aj,
  127, 1747

\bibitem[{{Hartmann}(2002)}]{Hartmann2002}
{Hartmann}, L. 2002, \apjl, 566, L29

\bibitem[{{Hartmann} {et~al.}(1986){Hartmann}, {Hewett}, {Stahler}, \&
  {Mathieu}}]{Hartmann1986}
{Hartmann}, L., {Hewett}, R., {Stahler}, S., \& {Mathieu}, R.~D. 1986, \apj,
  309, 275

\bibitem[{{Hartmann} \& {Kenyon}(1996)}]{Hartmann1996}
{Hartmann}, L. \& {Kenyon}, S.~J. 1996, \araa, 34, 207

\bibitem[{{Hartmann} \& {Stauffer}(1989)}]{Hartmann1989}
{Hartmann}, L. \& {Stauffer}, J.~R. 1989, \aj, 97, 873

\bibitem[{{Hauschildt} {et~al.}(1999){Hauschildt}, {Allard}, \&
  {Baron}}]{Hauschildt1999}
{Hauschildt}, P.~H., {Allard}, F., \& {Baron}, E. 1999, \apj, 512, 377

\bibitem[{{Herbst} {et~al.}(2001){Herbst}, {Bailer-Jones}, \&
  {Mundt}}]{Herbst2001}
{Herbst}, W., {Bailer-Jones}, C.~A.~L., \& {Mundt}, R. 2001, \apjl, 554, L197

\bibitem[{{Herbst} {et~al.}(2000){Herbst}, {Rhode}, {Hillenbrand}, \&
  {Curran}}]{Herbst2000}
{Herbst}, W., {Rhode}, K.~L., {Hillenbrand}, L.~A., \& {Curran}, G. 2000, \aj,
  119, 261

\bibitem[{{Jayawardhana} {et~al.}(2003){Jayawardhana}, {Mohanty}, \&
  {Basri}}]{Jayawardhana2003a}
{Jayawardhana}, R., {Mohanty}, S., \& {Basri}, G. 2003, \apj, 592, 282

\bibitem[{{Johns-Krull} \& {Gafford}(2002)}]{Johns-Krull2002}
{Johns-Krull}, C.~M. \& {Gafford}, A.~D. 2002, \apj, 573, 685

\bibitem[{{Johns-Krull} {et~al.}(1999){Johns-Krull}, {Valenti}, {Hatzes}, \&
  {Kanaan}}]{Johns-Krull1999}
{Johns-Krull}, C.~M., {Valenti}, J.~A., {Hatzes}, A.~P., \& {Kanaan}, A. 1999,
  \apjl, 510, L41

\bibitem[{{Jones} {et~al.}(1996){Jones}, {Fischer}, \& {Stauffer}}]{Jones1996}
{Jones}, B.~F., {Fischer}, D.~A., \& {Stauffer}, J.~R. 1996, \aj, 112, 1562

\bibitem[{{Kaas} {et~al.}(2004){Kaas}, {Olofsson}, {Bontemps}, {Andr{\' e}},
  {Nordh}, {Huldtgren}, {Prusti}, {Persi}, {Delgado}, {Motte}, {Abergel},
  {Boulanger}, {Burgdorf}, {Casali}, {Cesarsky}, {Davies}, {Falgarone},
  {Montmerle}, {Perault}, {Puget}, \& {Sibille}}]{Kaas2004}
{Kaas}, A.~A., {Olofsson}, G., {Bontemps}, S., {Andr{\' e}}, P., {Nordh}, L.,
  {Huldtgren}, M., {Prusti}, T., {Persi}, P., {Delgado}, A.~J., {Motte}, F.,
  {Abergel}, A., {Boulanger}, F., {Burgdorf}, M., {Casali}, M.~M., {Cesarsky},
  C.~J., {Davies}, J., {Falgarone}, E., {Montmerle}, T., {Perault}, M.,
  {Puget}, J.~L., \& {Sibille}, F. 2004, \aap, 421, 623

\bibitem[{{Kenyon} \& {Hartmann}(1995)}]{Kenyon1995}
{Kenyon}, S.~J. \& {Hartmann}, L. 1995, \apjs, 101, 117

\bibitem[{{Kenyon} {et~al.}(1990){Kenyon}, {Hartmann}, {Strom}, \&
  {Strom}}]{Kenyon1990}
{Kenyon}, S.~J., {Hartmann}, L.~W., {Strom}, K.~M., \& {Strom}, S.~E. 1990,
  \aj, 99, 869

\bibitem[{{K\"onigl}(1991)}]{Konigl1991}
{K\"onigl}, A. 1991, \apjl, 370, L39

\bibitem[{{Luhman} \& {Rieke}(1999)}]{Luhman1999}
{Luhman}, K.~L. \& {Rieke}, G.~H. 1999, \apj, 525, 440

\bibitem[{{Magazzu} {et~al.}(1992){Magazzu}, {Rebolo}, \&
  {Pavlenko}}]{Magazzu1992}
{Magazzu}, A., {Rebolo}, R., \& {Pavlenko}, I.~V. 1992, \apj, 392, 159

\bibitem[{{Mart{\'{\i}}n} {et~al.}(2000){Mart{\'{\i}}n}, {Koresko}, {Kulkarni},
  {Lane}, \& {Wizinowich}}]{Martin2000}
{Mart{\'{\i}}n}, E.~L., {Koresko}, C.~D., {Kulkarni}, S.~R., {Lane}, B.~F., \&
  {Wizinowich}, P.~L. 2000, \apjl, 529, L37

\bibitem[{{Massey}(1997)}]{Massey1997}
{Massey}, P. 1997, {A User's Guide to CCD Reductions with IRAF }, {National
  Optical Astronomy Observatory}

\bibitem[{{Massey} {et~al.}(1992){Massey}, {Valdes}, \& {Barnes}}]{Massey1992}
{Massey}, P., {Valdes}, F., \& {Barnes}, J. 1992, {A User's Guide to Reducing
  Slit Spectra with IRAF}, {National Optical Astronomy Observatory}

\bibitem[{{Mathieu}(2003)}]{Mathieu2003}
{Mathieu}, R. 2003, in Proc. of the IAU Symposium 215, A. Maeder and P. Eenens;
  Ed., in press (see astro-ph 0303199)

\bibitem[{{Matt} \& {Pudritz}(2004)}]{Matt2004}
{Matt}, S. \& {Pudritz}, R.~E. 2004, \apjl, 607, L43

\bibitem[{{McLean} {et~al.}(1998){McLean}, {Becklin}, {Bendiksen}, {Brims},
  {Canfield}, {Figer}, {Graham}, {Hare}, {Lacayanga}, {Larkin}, {Larson},
  {Levenson}, {Magnone}, {Teplitz}, \& {Wong}}]{mclean1998}
{McLean}, I.~S., {Becklin}, E.~E., {Bendiksen}, O., {Brims}, G., {Canfield},
  J., {Figer}, D.~F., {Graham}, J.~R., {Hare}, J., {Lacayanga}, F., {Larkin},
  J.~E., {Larson}, S.~B., {Levenson}, N., {Magnone}, N., {Teplitz}, H., \&
  {Wong}, W. 1998, in Proc. SPIE Vol. 3354, p. 566-578, Infrared Astronomical
  Instrumentation, Albert M. Fowler; Ed., 566--578

\bibitem[{{Mohanty} \& {Basri}(2003)}]{Mohanty2003}
{Mohanty}, S. \& {Basri}, G. 2003, \apj, 583, 451

\bibitem[{{Muzerolle} {et~al.}(1998){Muzerolle}, {Hartmann}, \&
  {Calvet}}]{Muzerolle1998}
{Muzerolle}, J., {Hartmann}, L., \& {Calvet}, N. 1998, \aj, 116, 2965

\bibitem[{{Neuhaeuser} {et~al.}(1995){Neuhaeuser}, {Sterzik}, {Schmitt},
  {Wichmann}, \& {Krautter}}]{Neuhaeuser1995}
{Neuhaeuser}, R., {Sterzik}, M.~F., {Schmitt}, J.~H.~M.~M., {Wichmann}, R., \&
  {Krautter}, J. 1995, \aap, 297, 391

\bibitem[{{Ochsenbein} {et~al.}(2000){Ochsenbein}, {Bauer}, \&
  {Marcout}}]{Ochsenbein2000}
{Ochsenbein}, F., {Bauer}, P., \& {Marcout}, J. 2000, \aaps, 143, 23

\bibitem[{{Padgett}(1996)}]{Padgett1996}
{Padgett}, D.~L. 1996, \apj, 471, 847

\bibitem[{{Popham}(1996)}]{Popham1996}
{Popham}, R. 1996, \apj, 467, 749

\bibitem[{{Preibisch} \& {Smith}(1997)}]{Preibisch1997}
{Preibisch}, T. \& {Smith}, M.~D. 1997, \aap, 322, 825

\bibitem[{{Press} {et~al.}(1986){Press}, {Flannery}, {Teukolsky}, \&
  {Vetterling}}]{Press1986}
{Press}, W.~H., {Flannery}, B.~P., {Teukolsky}, S.~A., \& {Vetterling}, W.~T.
  1986, Numerical recipes: the art of scientific computing (Cambridge
  University Press)

\bibitem[{{Prosser}(1992)}]{Prosser1992}
{Prosser}, C.~F. 1992, \aj, 103, 488

\bibitem[{{Prosser}(1994)}]{Prosser1994}
---. 1994, \aj, 107, 1422

\bibitem[{{Queloz} {et~al.}(1998){Queloz}, {Allain}, {Mermilliod}, {Bouvier},
  \& {Mayor}}]{Queloz1998}
{Queloz}, D., {Allain}, S., {Mermilliod}, J.-C., {Bouvier}, J., \& {Mayor}, M.
  1998, \aap, 335, 183

\bibitem[{{Rebull}(2001)}]{Rebull2001}
{Rebull}, L.~M. 2001, \aj, 121, 1676

\bibitem[{{Rebull} {et~al.}(2004){Rebull}, {Wolff}, \& {Strom}}]{Rebull2004}
{Rebull}, L.~M., {Wolff}, S.~C., \& {Strom}, S.~E. 2004, \aj, 127, 1029

\bibitem[{{Rhode} {et~al.}(2001){Rhode}, {Herbst}, \& {Mathieu}}]{Rhode2001}
{Rhode}, K.~L., {Herbst}, W., \& {Mathieu}, R.~D. 2001, \aj, 122, 3258

\bibitem[{{Shu} {et~al.}(1994){Shu}, {Najita}, {Ostriker}, {Wilkin}, {Ruden},
  \& {Lizano}}]{Shu1994}
{Shu}, F., {Najita}, J., {Ostriker}, E., {Wilkin}, F., {Ruden}, S., \&
  {Lizano}, S. 1994, \apj, 429, 781

\bibitem[{{Sicilia-Aguilar} {et~al.}(2005){Sicilia-Aguilar}, {Hartmann},
  {Szentgyorgyi}, {Fabricant}, {F{\H u}r{\' e}sz}, {Roll}, {Conroy}, {Calvet},
  {Tokarz}, \& {Hern{\' a}ndez}}]{Sicilia-Aguilar2005}
{Sicilia-Aguilar}, A., {Hartmann}, L.~W., {Szentgyorgyi}, A.~H., {Fabricant},
  D.~G., {F{\H u}r{\' e}sz}, G., {Roll}, J., {Conroy}, M.~A., {Calvet}, N.,
  {Tokarz}, S., \& {Hern{\' a}ndez}, J. 2005, \aj, 129, 363

\bibitem[{{Simon} {et~al.}(1995){Simon}, {Ghez}, {Leinert}, {Cassar}, {Chen},
  {Howell}, {Jameson}, {Matthews}, {Neugebauer}, \& {Richichi}}]{Simon1995}
{Simon}, M., {Ghez}, A.~M., {Leinert}, C., {Cassar}, L., {Chen}, W.~P.,
  {Howell}, R.~R., {Jameson}, R.~F., {Matthews}, K., {Neugebauer}, G., \&
  {Richichi}, A. 1995, \apj, 443, 625

\bibitem[{{Sneden}(1973)}]{Sneden1973}
{Sneden}, C.~A. 1973, Ph.D.~Thesis

\bibitem[{{Soderblom} {et~al.}(1993){Soderblom}, {Stauffer}, {Hudon}, \&
  {Jones}}]{Soderblom1993}
{Soderblom}, D.~R., {Stauffer}, J.~R., {Hudon}, J.~D., \& {Jones}, B.~F. 1993,
  \apjs, 85, 315

\bibitem[{{Stassun} {et~al.}(1999){Stassun}, {Mathieu}, {Mazeh}, \&
  {Vrba}}]{Stassun1999}
{Stassun}, K.~G., {Mathieu}, R.~D., {Mazeh}, T., \& {Vrba}, F.~J. 1999, \aj,
  117, 2941

\bibitem[{{Stassun} {et~al.}(2001){Stassun}, {Mathieu}, {Vrba}, {Mazeh}, \&
  {Henden}}]{Stassun2001}
{Stassun}, K.~G., {Mathieu}, R.~D., {Vrba}, F.~J., {Mazeh}, T., \& {Henden}, A.
  2001, \aj, 121, 1003

\bibitem[{{Stauffer} {et~al.}(1984){Stauffer}, {Hartmann}, {Soderblom}, \&
  {Burnham}}]{Stauffer1984}
{Stauffer}, J.~R., {Hartmann}, L., {Soderblom}, D.~R., \& {Burnham}, N. 1984,
  \apj, 280, 202

\bibitem[{{Stauffer} \& {Hartmann}(1987)}]{Stauffer1987}
{Stauffer}, J.~R. \& {Hartmann}, L.~W. 1987, \apj, 318, 337

\bibitem[{{Stauffer} {et~al.}(1989){Stauffer}, {Hartmann}, \&
  {Jones}}]{Stauffer1989}
{Stauffer}, J.~R., {Hartmann}, L.~W., \& {Jones}, B.~F. 1989, \apj, 346, 160

\bibitem[{{Stauffer} {et~al.}(1997){Stauffer}, {Hartmann}, {Prosser},
  {Randich}, {Balachandran}, {Patten}, {Simon}, \& {Giampapa}}]{Stauffer1997}
{Stauffer}, J.~R., {Hartmann}, L.~W., {Prosser}, C.~F., {Randich}, S.,
  {Balachandran}, S., {Patten}, B.~M., {Simon}, T., \& {Giampapa}, M. 1997,
  \apj, 479, 776

\bibitem[{{Tout} \& {Pringle}(1992)}]{Tout1992}
{Tout}, C.~A. \& {Pringle}, J.~E. 1992, \mnras, 256, 269

\bibitem[{{Walter} {et~al.}(1988){Walter}, {Brown}, {Mathieu}, {Myers}, \&
  {Vrba}}]{Walter1988}
{Walter}, F.~M., {Brown}, A., {Mathieu}, R.~D., {Myers}, P.~C., \& {Vrba},
  F.~J. 1988, \aj, 96, 297

\bibitem[{{White} \& {Hillenbrand}(2004)}]{White2004}
{White}, R. \& {Hillenbrand}, L. 2004, ArXiv Astrophysics e-prints

\bibitem[{{White} \& {Basri}(2003)}]{White2003}
{White}, R.~J. \& {Basri}, G. 2003, \apj, 582, 1109

\bibitem[{{White} {et~al.}(1999){White}, {Ghez}, {Reid}, \&
  {Schultz}}]{White1999}
{White}, R.~J., {Ghez}, A.~M., {Reid}, I.~N., \& {Schultz}, G. 1999, \apj, 520,
  811

\bibitem[{{Wichmann} {et~al.}(2000){Wichmann}, {Torres}, {Melo}, {Frink},
  {Allain}, {Bouvier}, {Krautter}, {Covino}, \& {Neuh{\"
  a}user}}]{Wichmann2000}
{Wichmann}, R., {Torres}, G., {Melo}, C.~H.~F., {Frink}, S., {Allain}, S.,
  {Bouvier}, J., {Krautter}, J., {Covino}, E., \& {Neuh{\" a}user}, R. 2000,
  \aap, 359, 181

\bibitem[{{Wilking} {et~al.}(1989){Wilking}, {Lada}, \& {Young}}]{Wilking1989}
{Wilking}, B.~A., {Lada}, C.~J., \& {Young}, E.~T. 1989, \apj, 340, 823

\end{thebibliography}
\end{document}